\documentclass[twocolumn,prd,floatfix,amsmath,nofootinbib,amssymb,floatfix]{revtex4}
\usepackage{graphicx,color,dcolumn,booktabs,bm}
\usepackage{longtable,lscape}
\usepackage{txfonts}
\usepackage{overpic}
\usepackage{amssymb}
\usepackage{indentfirst}
\usepackage{feynmf}
\usepackage{slashed}
\usepackage{cases}
\usepackage{multirow}
\usepackage[colorlinks,
            citecolor=blue,
            anchorcolor=red,
            menucolor=red,
            linkcolor=red,
            filecolor=red,
            runcolor=red,
            urlcolor=blue,
            frenchlinks=red]{hyperref}
\usepackage{epstopdf}
\usepackage{bm}
\begin{document}
\title{The role of the short-distance interaction in $e^+e^-\to \gamma X(3872)$}
\author{Ming-Xiao Duan$^{1}$}\email{duanmx@ihep.ac.cn}
\affiliation{
$^1$ Institute of High Energy Physics, Chinese Academy of Sciences, Beijing 100049, People's Republic of China}

\begin{abstract}
In this study, we analyze the cross-section data from the $e^+e^-\to \gamma J/\psi\omega$ process to explore both short-distance and long-distance interactions for the radiative transition $Y(4200)\to \gamma X(3872)$. We investigate the short-distance effects through the E1 transition among the $c\bar{c}$ components, and the long-distance effects via hadronic loop diagrams. Our numerical analysis reveals that short-distance interactions play a significantly larger role in the radiative transition than that of the long-distance interactions. This finding underscores the importance of the compact $c\bar{c}$ components in both the initial and final states for accurately understanding the cross-section $\sigma[e^+e^-\to \gamma J/\psi\omega]$. Furthermore, with the help of relative branch ratio $\mathcal{R}$ we estimate the $\Gamma[Y(4200)\to\gamma X(3872)]$ implied in the the experimental study. Finally, we also discuss the possible existence of $\psi(4040)$ signal within the cross-section data.
\end{abstract}

\maketitle

\section{introduction}\label{sec1}
As early as 2014, the BESIII collaboration analyzed the $e^+e^-\to \gamma J/\psi\pi\pi$ process, revealing a resonant peak around 4.2 GeV and the $X(3872)$ in its radiative decay process \cite{BESIII:2013fnz}. This measurement was the first observation of the $X(3872)$ in a radiative decay process within the $J/\psi\pi\pi$ invariant mass spectrum, which offered an unprecedented opportunity to probe the internal structure of the $X(3872)$. Subsequently, a more comprehensive analysis conducted in 2019 with an expanded dataset \cite{BESIII:2019qvy} studied the process $e^+e^-\to Y(4200)\to \gamma X(3872)\to \gamma\omega J/\psi(\gamma J/\psi \pi\pi)$ across the energy range \(4.0076~{\rm GeV} < \sqrt{s} < 4.5995~{\rm GeV}\). This analysis identified a resonant peak attributed to $Y(4200)$ which characterized by a mass $M_{Y(4200)} = 4200.6^{+7.9}_{-13.3} \pm 3.0~{\rm MeV}$ and a width $\Gamma_{Y(4200)} = 115^{+38}_{-26} \pm 12~{\rm MeV}$. The \(X(3872)\) in the \(J/\psi\omega (J/\psi\pi\pi)\) invariant mass spectrum is outstanding in the radiative decay of $Y(4200)$, which further emphasize $e^+e^-\to \gamma J/\psi\omega$ process is important in understanding exotic hadron structures.

In addition to the radiative process, the $X(3872)$ has been observed in a variety of production channels \cite{Belle:2003nnu, LHCb:2013kgk, Belle:2006olv, BaBar:2004oro, CMS:2013fpt, LHCb:2020xds, CDF:2003cab, D0:2004zmu, BESIII:2013fnz, BESIII:2019qvy, BESIII:2022bse, COMPASS:2017wql}, with its mass and width accurately measured at $M_{X(3872)} = 3871.65\pm0.06~{\rm MeV}$ and $\Gamma_{X(3872)} = 1.19\pm0.21~{\rm MeV}$, respectively \cite{ParticleDataGroup:2022pth}. This positions the $X(3872)$ as a shallow bound state, with a mass notably below the predicted value for the \(\chi_{c1}(2P)\) charmonium state and very close to the threshold of the $D\bar{D}^*$ channel \cite{Barnes:2003vb, Ortega:2009hj}. Such observations challenge the identification of $X(3872)$ as a pure charmonium state, but support the hypothesis of a $D\bar{D}^*$ molecular structure \cite{Wong:2003xk, AlFiky:2005jd, Swanson:2003tb}. According to the Weinberg's compositeness theorem, the $X(3872)$ could be a hybrid of loosely bound \(D\bar{D}^*\) pairs and a compact \(c\bar{c}\) core. This compositeness scheme of the $X(3872)$ is supported by the significant production rates of $X(3872)$ in $B$ meson decays and $pp$ collisions \cite{Weinberg:1962hj, Weinberg:1965zz, Suzuki:2005ha}. Theoretical studies, including coupled channel analyses and lattice QCD calculations, suggest that to understand the experimental data of the $X(3872)$, it is necessary to consider both the $c\bar{c}$ and $D\bar{D}^*$ components \cite{Barnes:2003vb, Ortega:2009hj, Kalashnikova:2005ui, Zhou:2013ada, Ono:1983rd, Danilkin:2010cc, Prelovsek:2013cra, Padmanath:2015era}. For example, an analysis of LHC data within the non-relativistic QCD framework estimates the $c\bar{c}$ core's contribution to the $X(3872)$ structure is from 28\% to 44\% \cite{Meng:2013gga}. Additionally, analysis of the $X(3872)$'s lineshape in the $D\bar{D}^*$ invariant mass spectrum from the $B \to KX(3872) \to KD\bar{D}^*$ process suggests a $c\bar{c}$ component of $0.19 \pm 0.29$ \cite{Chen:2013upa}. These findings reveal the significant role of $c\bar{c}$ component in understanding the nature of the $X(3872)$.

\begin{figure}[h!]
\begin{center}
\includegraphics[width=7.9cm,keepaspectratio]{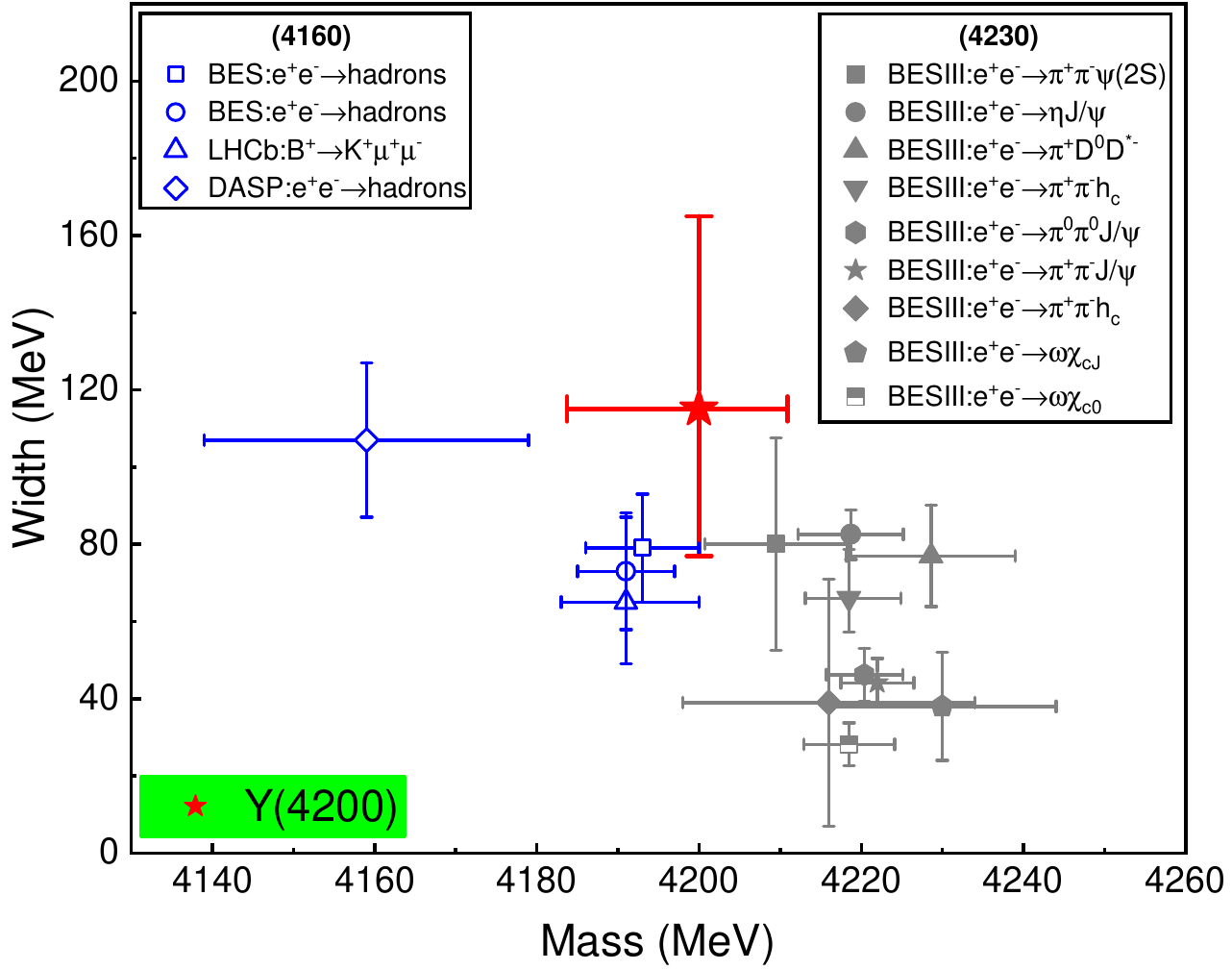}
\caption{The comparisons of the measured masses and widths of $\psi(4160)$ and $\psi(4230)$ across different channels with $Y(4200)$.}
\label{ferror}
\end{center}
\end{figure}

Following the measurement of $X(3872)$ in $e^+e^- \to \gamma X(3872)$ in 2013, theorists analyzed the radiative decay width of $Y(4260) \to X(3872)\gamma$ using hadronic loops within the compositeness framework, and can explain the data available at that time \cite{Guo:2013zbw, Dong:2014zka}. Nonetheless, the emergence of new experimental data in 2019, including the refined resonance parameters for the vector structure $Y(4200)$ and the cross-section measurements for $e^+e^- \to \gamma J/\psi\omega$, supports an updated examination of the radiative decay process to accommodate these latest experimental data.

In the latest experimental datasets, the $Y(4200)$ structure is observed to have a relatively wide decay width according to experimental fits, which position it close to two vector charmonium-like states $\psi(4230)$ and $\psi(4160)$ \cite{ParticleDataGroup:2022pth, vonDetten:2024eie}. This proximity introduces the possibility that either or both of these charmonium-like states could be the original source of the $Y(4200)$ structure. To clearly depict this scenario, we present a comparison of the resonance parameters of $\psi(4160)$, $\psi(4230)$ and $Y(4200)$ in Fig.~\ref{ferror} \cite{DASP:1978dns, Mo:2010bw, BES:2007zwq, LHCb:2013ywr, Yuan:2013uta, BESIII:2018iea, BESIII:2016bnd, BESIII:2016adj, BESIII:2017tqk, BESIII:2014rja, Ablikim:2020pzw, BESIII:2020bgb, BESIII:2019gjc}.

When comparing the experimental data plotted in Fig.~\ref{ferror}, it becomes apparent that the $Y(4200)$ structure cannot be definitively identified as either the $\psi(4160)$ or $\psi(4230)$ state with the current dataset. Consequently, this study will explore both possibilities. In the framework of potential models, $\psi(4160)$ and $\psi(4230)$ are recognized as the $\psi(2D)$ and $\psi(4S)$ charmonium states, respectively \cite{He:2014xna, Wang:2019mhs, Li:2009zu}. This classification indicates the presence of a compact $c\bar{c}$ component in the initial state of the $Y(4200) \to \gamma X(3872)$ process. Given the existence of $c\bar{c}$ components in both the initial $Y(4200)$ structure and final $X(3872)$ state, the E1 transition becomes a crucial process, especially it is in the leading order comparing with the long-distance interaction. Therefore, our analysis incorporates the E1 transition from the $\psi(2D)$ component within the $Y(4200)$ structure to the $\chi_{c1}(2P)$ component within the $X(3872)$. Additionally, we also calculate the radiative transition involving long-distance interactions. Through this comprehensive analysis, we aim to present the decay width of $Y(4200) \to \gamma X(3872)$ alongside the numerical cross-section $\sigma[e^+e^- \to \gamma\omega J/\psi]$, trying to understand the corresponding experimental data.

The paper is organized with four parts. Following the introduction, the formalism is given in Sec.~\ref{sec2} to illustrate the numerical calculations for the radiative transition in the short-distance and long-distance interaction. Then, we discuss the numerical results in the Sec.~\ref{sec3}. Finally, a summary in Sec.~\ref{sec4} is shown at the end of this work.

\section{Formalism}\label{sec2}
\begin{center}
\begin{figure}[htbp]
\includegraphics[width=8.2cm,keepaspectratio]{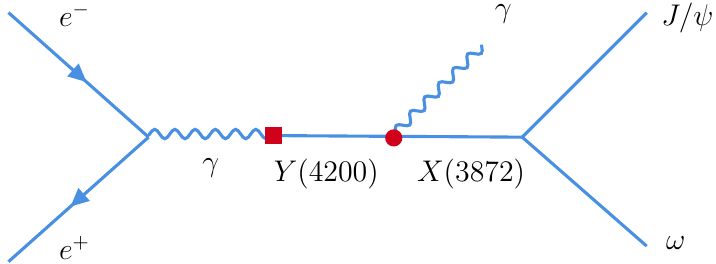}
\caption{The Feynman diagram of the $e^+e^-\to\gamma X(3872)\to \gamma J/\psi\omega$ scattering process.}
\label{f1}
\end{figure}
\end{center}

The Feynman diagram of the scattering process $e^+e^-\to \gamma J/\psi\omega$ \cite{BESIII:2019qvy} is depicted in Fig.~\ref{f1}. This diagram includes the radiative transition $Y(4200)\to \gamma X(3872)$. As discussed in the introduction, current cross-section data are insufficient to determine whether the vector structure $Y(4200)$ originates from $\psi(4160)$ or $\psi(4230)$.  Therefore, we will discuss the both scenarios in the subsequent sections.

In Fig.~\ref{f1}, the coupling between intermediate photon and the vector structure $Y(4200)$ is expressed by the vector meson dominance (VMD) vertex \cite{Kroll:1967it}.  The Lagrangian of VMD vertex is
\begin{equation}
\mathcal{L}_{\gamma Y}=-\frac{eM_Y^2}{f_Y}V^\mu A_\mu,
\end{equation}
where the $V^\mu$ and $A_\mu$ are the fields of the vector meson and photon. $M_Y$ and $f_Y$ are the mass and decay constant of the vector state, respectively. In the calculation, $f_{Y_1}=44$ and $f_{Y_2}=24$ \footnote{The subscripts $Y_{1}$ and $Y_2$ correspond to $\psi(4160)$ and $\psi(4230)$, respectively.} are fixed by the central value of $\Gamma[\psi(4160)\to e^+e^-]=0.48\pm0.22~{\rm keV}$ and $\Gamma[\psi(4230)\to e^+e^-]\sim\Gamma[\psi(4230)\to \mu^+\mu^-]=1.53\pm1.26\pm0.54~{\rm keV}$, respectively \cite{ParticleDataGroup:2022pth}.
For the interaction of $X(3872)$ with $J/\psi \omega$, the vertex is given by the effective Lagrangian
\begin{equation}
\mathcal{L}=g_{XJ/\psi\omega}\epsilon^{\mu\nu\alpha\beta}\partial_\mu X_\nu \psi_\alpha \omega_\beta,
\end{equation}
where $X_\nu$, $\psi_\alpha$, and $\omega_\beta$ are the fields of $X(3872)$, $J/\psi$, and $\omega$, respectively. The coupling constant $g_{XJ/\psi\omega}=0.135$ can be determined by the experimental data directly. The remaining unknown coupling between $Y(4200)$ and $\gamma X(3872)$ in Fig.~\ref{f1} can be described by the Lagrangian, i.e.,
\begin{equation}
\mathcal{L}=ieg_{YX\gamma}\varepsilon_{\mu\nu\alpha\beta}F^{\mu\nu}Y^\alpha X^\beta.
\end{equation}

Taking into account the long-distance $D\bar{D}^*$ component and the short-distance $\chi_{c1}(2P)$ component, the $X(3872)$ can be written as
\begin{equation}\label{eq3872}
\begin{split}
|X(3872)\rangle=\sqrt{Z_X}|\chi_{c1}(2P)\rangle+\sqrt{1-Z_X}|D\bar{D}^*\rangle.
\end{split}
\end{equation}
Then, the physical propagator of $X(3872)$ is \cite{Chen:2013upa}
\begin{equation}
\mathcal{P}(E)=\frac{iZ_X}{E+B+\Sigma(E,Z_X)+i\frac{\Gamma_X}{2}},
\end{equation}
where the propagator is given in the nonrelativistic limit, since the binding energy of $X(3872)$ to $D\bar{D}^*$ channel is very small. $\Sigma(E,Z_X)$ is the self-energy function of $X(3872)$, which is expressed as \cite{Chen:2013upa}
\begin{equation}
\Sigma(E,Z_X)=-g_X^2[\frac{\mu}{2\pi}\sqrt{-2\mu E-i\epsilon}+\frac{\mu\sqrt{2\mu B}}{4\pi B}(E-B)].
\end{equation}
In the equation, $\mu$ represents the reduced mass of the $D\bar{D}^*$ pairs, and the binding energy $B$ is determined by $B=m_D+m_{D^*}-m_{X(3872)}$. The variable $E$ denotes the energy of the $X(3872)$ relative to the threshold of the $D\bar{D}^*$ channel.

With the above preparation, the total amplitude of $e^+e^-\to\gamma J/\psi\omega$ can be written as
\begin{equation}\label{eqamptot}
\begin{split}
i\mathcal{M}^{\rm tot}=&\bar{v}(p_2)(-ie\gamma^\mu)u(p_1)\frac{-ig_{\mu\nu}}{q^2}(-i\frac{eM_V^2}{f_V})\frac{i(-g^{\nu\phi}+\frac{q^\nu q^\phi}{m_Y^2})}{q^2-m_Y^2+i\sqrt{s}\Gamma_Y}\\
&\times i(\sqrt{Z_X}g^S_{YX\gamma}+\sqrt{1-Z_X}\frac{g_{YX\gamma}^L}{\sqrt{Z_X}})\varepsilon_{\eta\tau\phi\omega}(ip_5^\eta\varepsilon_5^{*\tau}-ip_5^\tau\varepsilon_5^{*\eta})\\
&\times \frac{i(-g^{\omega\lambda}+\frac{k^\omega k^\lambda}{m_X^2})Z_X}{E+B+\Sigma(E,Z_X)+i\frac{\Gamma_X}{2}}ig_{XJ\psi\omega}\varepsilon_{\alpha\lambda\beta\delta}(-ik^\alpha)\varepsilon_3^{*\beta}\varepsilon_4^{*\delta},\\
\end{split}
\end{equation}
where $g_{YX\gamma}^S$ and $g_{YX\gamma}^L$ denote the coupling constants for the short-distance interaction among $c\bar{c}$ components and the long-distance interaction resulting from the $D\bar{D}^*$ component of $X(3872)$, respectively. To estimate $g_{YX\gamma}^S$, we compute the E1 transition between the charmonium components from the initial $Y(4200)$ structure to the final $X(3872)$ state. Meanwhile, $g_{YX\gamma}^L$ is derived from the decay width through detailed calculations of hadronic loop diagrams. These calculations are presented in detail in the following subsections.

\subsection{The E1 transition to the $c\bar{c}$ component in $X(3872)$}
In this subsection, we examine the radiative production of $X(3872)$ via short-distance interactions. In the charmonium family, the vector states $3^3S_1$, $2^3D_1$, $4^3S_1$, and $3^3D_1$ are approximately around 4.2 GeV, which is close to the $Y(4200)$. Hence, we calculate the E1 radiative decay from these states to $\gamma X(3872)$. The decay width, denoted as $\Gamma_{\rm E1}[\psi(mS)/\psi(nD)\to \gamma\chi_{c1}(2P)]$, is determined using the following formula \cite{Eichten:1979ms, Kwong:1988ae, Barnes:2005pb, Li:2012vc}, i.e.,
\begin{equation}
\begin{split}
&\Gamma_{\rm E1}\left(n^{2S+1}L_J\to n^{\prime 2S^\prime+1}L^\prime_{J^\prime}+\gamma\right)\\
&=\frac{4}{3}C_{fi}\delta_{SS^\prime}e_c^2\alpha|\langle f|r|i \rangle|^2E_\gamma^3,\\
\end{split}
\end{equation}
where the charge of charm quark and fine structure constant are $e_c=+\frac{2}{3}$ and $\alpha=\frac{1}{137}$, respectively. $n$, $L$, $S$, and $J$ ($n^\prime$, $L^\prime$, $S^\prime$, and $J^\prime$) are the radial quantum number, orbital angular momentum, spin, and total angular momentum of initial states (final states). $E_\gamma$ is the energy of the final photon. $C_{fi}$ is an matrix element corresponding to the angular momentum, which is defined as
\begin{equation}
\begin{split}
C_{fi}=max(L,L^\prime)(2J^\prime+1)\begin{Bmatrix}L^\prime & J^\prime & S \\ J & L & 1\end{Bmatrix}^2. \\
\end{split}
\end{equation}
The $\langle f|r|i \rangle$ is an overlap of radial wave functions of initial and final charmonium state, which is
\begin{equation}
\begin{split}
\langle f|r|i \rangle=\int_0^\infty rR_f(r)R_i(r)r^2{\rm d}r. \\
\end{split}
\end{equation}
In the overlap, $R(r)$ represents the radial wave functions of charmonium states, which can be obtained from the potential model in Ref.~\cite{Duan:2020tsx}. We calculate the $E1$ transition and list the corresponding results in Table~\ref{Tab1}.
\begin{table}[htbp]
\caption{The decay widths of E1 transitions in $\psi(mS)/\psi(nD)\to \gamma\chi_{c1}(2P)$ processes. $M_{\psi(4040)}=4.040~{\rm GeV}$ is employed as the mass for initial $\psi(3S)$ states. $\Gamma^1[\psi(2D)\to\gamma\chi_{c1}(2P)]$ and $\Gamma^2[\psi(2D)\to\gamma\chi_{c1}(2P)]$ are obtained by employing $M_{\psi(4160)}=4.191~{\rm GeV}$ and $M_{\psi(4230)}=4.222~{\rm GeV}$ as the initial mass of $\psi(2D)$ state, respectively \cite{ParticleDataGroup:2022pth}. Meanwhile, the $M_{\psi(4230)}=4.222$ GeV is employed to estimate the mass of the initial $\psi(4S/3D)$ state in the calculation.}
\label{Tab1}
\renewcommand\arraystretch{1.50}
\begin{tabular*}{86mm}{@{\extracolsep{20mm}}cc}%
\toprule[1.5pt]
\toprule[1pt]
Processes & Decay width (keV)\\
\midrule[1pt]
$\Gamma[\psi(3S)\to\gamma \chi_{c1}(2P)]$ &62\\
$\Gamma^1[\psi(2D)\to\gamma \chi_{c1}(2P)]$ &262\\
$\Gamma^2[\psi(2D)\to\gamma \chi_{c1}(2P)]$ &342\\
$\Gamma[\psi(4S)\to\gamma \chi_{c1}(2P)]$ &0.08\\
$\Gamma[\psi(3D)\to\gamma \chi_{c1}(2P)]$ &0.43\\
\bottomrule[1pt]
\bottomrule[1.5pt]
\end{tabular*}
\end{table}

With the above results, we find that the $\Gamma^{1(2)}[\psi(2D)\to\gamma \chi_{c1}(2P)]=262/342~{\rm keV}$ is significantly larger than the other channels in Table.~\ref{Tab1}, which suggests the $\psi(2D)\to\gamma\psi(2P)$ process dominates the short-distance interaction among the different possible E1 transitions.
Consequently, the short-distance coupling constant is determined as $g_{YX\gamma}^S$=2.75 by the $\Gamma^{1(2)}[\psi(2D)\to\gamma \chi_{c1}(2P)]$. \footnote{The different decay widths between $\Gamma^{1}[\psi(2D)\to\gamma \chi_{c1}(2P)]$ and $\Gamma^{2}[\psi(2D)\to\gamma \chi_{c1}(2P)]$ come from the different phase spaces of the initial $\psi(4160)$ and $\psi(4230)$, as a result the short-distance coupling constant $g_{YX\gamma}^S$=2.75 is found to be same.}

\subsection{Radiative production of $X(3872)$ through the long-distance contributions}

\begin{center}
\begin{figure}[htbp]
\includegraphics[width=8.0cm,keepaspectratio]{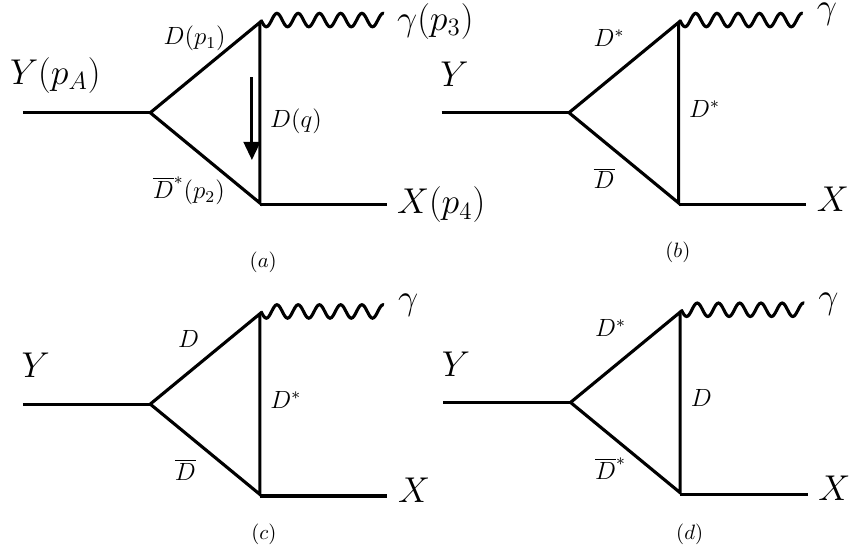}
\caption{The long-distance decay of $Y(4200)\to \gamma X(3872)$ through charmed meson loops.}
\label{f2}
\end{figure}
\end{center}

Since the molecular $D\bar{D}^*$ component is included in the $X(3872)$, $Y(4200)\to \gamma X(3872)$ process can take place through the meson loop diagrams, which represents the long-range interaction in radiative decay process. The Feynman diagram of the long-range process is shown in Fig.~\ref{f2}.

In the process, the vector structure $Y(4200)$ couples with a pair of $D^{(*)}\bar{D}^{(*)}$, then the $D^{(*)}\bar{D}^{(*)}$ pair scatters into a photon $\gamma$ and an $X(3872)$. The vertexes depicting the vector state $Y$ coupling with $D^{(*)}\bar{D}^{(*)}$ in Fig.~\ref{f2} are determined by the effective Lagrangian \cite{Wise:1992hn, Colangelo:2003sa, Casalbuoni:1996pg}, i.e.,
\begin{equation}
\begin{split}
&\mathcal{L}_{YD^{(*)}\bar{D}^{(*)}}\\
&=+ig_{YD\bar{D}}D\overleftrightarrow{\partial}^\alpha D^\dag Y_\alpha\\
&\quad+g_{YD\bar{D}^*}\varepsilon^{\mu\nu\alpha\beta}[D_\mu^{*}\overleftrightarrow{\partial_\nu}D^\dag-
D\overleftrightarrow{\partial_\nu}D_\mu^{*\dag}]\partial_\beta Y_\alpha\\
&\quad+ig_{YD^*\bar{D}^*}Y_\alpha[D^{*\alpha}\overleftrightarrow{\partial}_\mu D^{*\mu\dag}+D^*_\mu\overleftrightarrow{\partial}^\mu D^{*\alpha\dag}-D_\mu^*\overleftrightarrow{\partial}^\alpha D^{*\mu\dag}],\\
\end{split}
\end{equation}
where the heavy quark symmetry is considered in the Lagrangians. From the $\Gamma^{\rm tot}_{\psi(4160)}=70~{\rm MeV}$, the coupling constants of the above Lagrangians can be determined as $g_{Y_1D\bar{D}}=1.66$, $g_{Y_1D\bar{D}^*}=0.41$, and $g_{Y_1D^*\bar{D}^*}=1.79$. If the $\Gamma^{\rm tot}_{\psi(4230)}=48~{\rm MeV}$ is employed, the coupling constants can be evaluated as $g_{Y_2D\bar{D}}=1.17$, $g_{Y_2D\bar{D}^*}=0.29$, and $g_{Y_2D^*\bar{D}^*}=1.27$.

The vertex of $\gamma$ and $D^{(*)}\bar{D}^{(*)}$ is expressed by the Lagrangians, i.e.,
\begin{equation}
\begin{split}
\mathcal{L}_{D^+D^+\gamma}=&ieA_\mu D^-\overleftrightarrow{\partial}^\mu D^+,\\
\mathcal{L}_{DD^{*}\gamma}=&\frac{eg_{D^+D^{*-}\gamma}}{4}\varepsilon^{\mu\nu\alpha\beta}F_{\mu\nu}D^{*+}_{\alpha\beta}D^-\\
&+\frac{eg_{D^0\bar{D}^{*0}\gamma}}{4}\varepsilon^{\mu\nu\alpha\beta}F_{\mu\nu}D^{*0}_{\alpha\beta}\bar{D}^0,\\
\mathcal{L}_{D^{*+}D^{*+}\gamma}=&ieA_\mu[g^{\alpha\beta}D_\alpha^{*-}\overleftrightarrow{\partial}^\mu D_\beta^{*+}+g^{\mu\beta}D_\alpha^{*-}\partial^\alpha D_\beta^{*+}\\
&-g^{\mu\alpha}\partial^\beta D_\alpha^{*-}D_\beta^{*+}],\\
\end{split}
\end{equation}
where $D_{\alpha\beta}^*$ represents $D^*_{\alpha\beta}=\partial_\alpha D_\beta^*-\partial_\beta D^*_\alpha$. $g_{D^0\bar{D}^{*0}\gamma}$ and $g_{D^+D^{*-}\gamma}$ are estimated as~\cite{Chen:2010re}
\begin{equation}
\begin{split}
g_{D^0\bar{D}^{*0}\gamma}=2.0~\rm{GeV}^{-1},~g_{D^+D^{*-}\gamma}=-0.5~\rm{GeV}^{-1}.\\
\end{split}
\end{equation}

The probability of finding the $D\bar{D}^*$ molecular state and a $c\bar{c}$ compact state in an $X(3872)$ are expressed as $1-Z_X$ and $Z_X$, then the coupling between $X(3872)$ and $D\bar{D}^*$ pair is expressed as~\cite{Chen:2013upa}
\begin{equation}\label{eq14}
g_X^2=\frac{2\pi\sqrt{2\mu B}}{\mu^2}(1-Z_X).
\end{equation}
In Eq.~(\ref{eq14}), $B$ is the value of binding energy and $\mu$ is reduced mass of $D$ and $\bar{D}^*$. The Lagrangian expressing the coupling between $X(3872)$ and $D\bar{D}^*$ pair is
\begin{equation}
\mathcal{L}_{XD\bar{D}^*}=\frac{1}{\sqrt{2}}g_{X}X_\mu(D^\dag D^{*\mu}-D^{*\mu\dag}D).
\end{equation}

With above preparations, the amplitudes of Fig.~\ref{f2} can be written as
\begin{equation}
\begin{split}
\mathcal{M}^a=&\int \frac{d^4q}{(2\pi)^4}(-i)g_{YD\bar{D}^*}\varepsilon^{\mu\nu\alpha\beta} p_{A\beta}\varepsilon_{A\alpha}\left(p_1-p_2\right)_\nu \\
&\times e\varepsilon_3^*\cdot\left(p_1+q\right)\frac{g_{X}}{\sqrt{2}}\varepsilon_4^{*\rho}\left(-g_{\mu\rho}+\frac{p_{2\mu}p_{2\rho}}{m_2^2}\right)\\
&\times \frac{\mathcal{F}^2(q^2)}{(p_1^2-m_1^2)(p_2^2-m_2^2)(q^2-m_q^2)},\\
\end{split}
\end{equation}

\begin{equation}
\begin{split}
\mathcal{M}^b=&\int \frac{d^4q}{(2\pi)^4}(-i)g_{YD\bar{D}^*}\varepsilon^{\mu\nu\alpha\beta} p_{A\beta}\varepsilon_{A\alpha}(p_1-p_2)_\nu \\
&\times e\varepsilon_{3\rho}^*[g^{\sigma\tau}(p_1+q)^\rho+g^{\rho\tau}p_1^\sigma+g^{\rho\sigma}q^\tau]\\
&\times\frac{g_{X}}{\sqrt{2}}\varepsilon_4^{*\delta}\left(-g_{\sigma\delta}+\frac{q_\sigma q_\delta}{m_q^2}\right)\left(-g_{\mu\tau}+\frac{p_{1\mu}p_{1\tau}}{m_1^2}\right)\\
&\times\frac{\mathcal{F}^2(q^2)}{(p_1^2-m_1^2)(p_2^2-m_2^2)(q^2-m_q^2)},\\
\end{split}
\end{equation}

\begin{equation}
\begin{split}
\mathcal{M}^c=&\int \frac{d^4q}{(2\pi)^4}ig_{YD\bar{D}}\varepsilon_A\cdot(p_1-p_2)
\frac{eg_{DD^*\gamma}}{4}\varepsilon^{\mu\nu\alpha\beta}\\
&\times\left(p_{3\mu}g_{\nu\sigma}-p_{3\nu}g_{\mu\sigma}\right)\left(q_\alpha g_{\beta\tau}-q_\beta g_{\alpha\tau}\right)\\
&\times\varepsilon_3^{*\sigma}\frac{g_{X}}{\sqrt{2}}\varepsilon_{4\rho}^*\left(-g^{\tau\rho}+\frac{q^\tau q^\rho}{m_q^2}\right)\\
&\times\frac{\mathcal{F}^2(q^2)}{(p_1^2-m_1^2)(p_2^2-m_2^2)(q^2-m_q^2)},\\
\end{split}
\end{equation}

\begin{equation}
\begin{split}
\mathcal{M}^d=&\int \frac{d^4q}{(2\pi)^4}ig_{YD^*\bar{D}^*}\left[(p_1-p_2)_\alpha g_{\beta\gamma}+(p_1-p_2)_\beta g_{\alpha\gamma}\right.\\
&\left.-(p_1-p_2)_\gamma g_{\alpha\beta}\right]\varepsilon_A^\gamma\frac{eg_{DD^*\gamma}}{4}\varepsilon^{\mu\nu\lambda\kappa}\left(p_{3\mu}g_{\nu\sigma}-p_{3\nu}g_{\mu\sigma}\right)\\
&\times\left(p_{1\lambda}g_{\kappa\tau}-p_{1\kappa}g_{\lambda\tau}\right)\frac{g_{X}}{\sqrt{2}}\varepsilon_3^\sigma\varepsilon_{4\rho}\left(-g^{\alpha\tau}+\frac{p_1^\alpha p_1^\tau}{m_1^2}\right)\\
&\times\left(-g^{\beta\rho}+\frac{p_2^\beta p_2^\rho}{m_2^2}\right)\frac{\mathcal{F}^2(q^2)}{(p_1^2-m_1^2)(q^2-m_q^2)(p_2^2-m_2^2)}.\\
\end{split}
\end{equation}

In the amplitudes, $\varepsilon_{A}$ represents the polarized vector of the initial $Y$ state. The form factor $\mathcal{F}(q^2)$ is introduced to involve the off-shell effects and the inner structures of the exchanged mesons and also to remove the UV divergences of the loop integrals of the triangle loop diagrams. In this work, the monopole from factor is employed \cite{Cheng:2004ru}
\begin{equation}\label{eqFF1}
\mathcal{F}(q^2)=\frac{m_q^2-\Lambda^2}{q^2-\Lambda^2},\\
\end{equation}
where $m_q$ and $q$ are the mass and momentum of exchanged charmed meson, respectively. The cutoff $\Lambda$ is parameterized as $\Lambda=m_q+\alpha\Lambda_{\rm QCD}$, where $\Lambda_{\rm QCD}=220$ MeV \cite{Cheng:2004ru}.

\begin{center}
\begin{figure}[htbp]
\includegraphics[width=8.0cm,keepaspectratio]{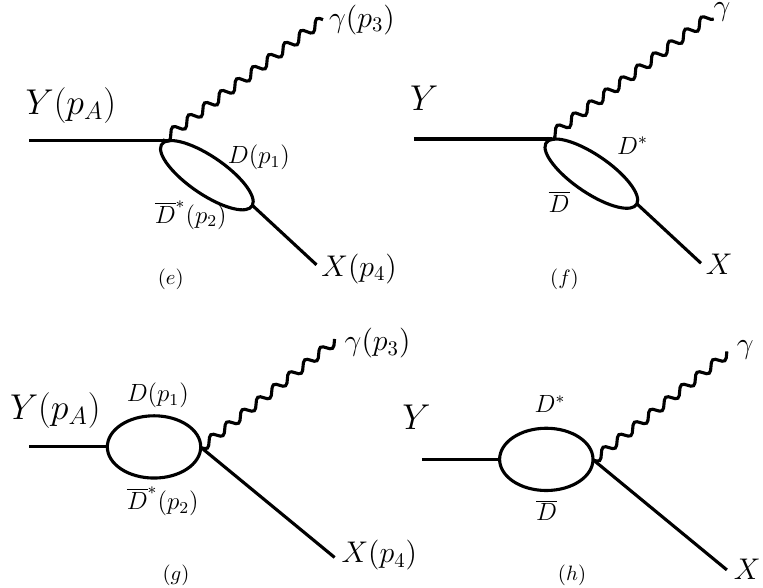}
\caption{The remaining contact terms for the $Y(4200)\to \gamma X(3872)$ process.}
\label{f4}
\end{figure}
\end{center}

We notice that the above long-distance processes depicted in Fig.~\ref{f2} can not ensure the gauge invariance of photon field. Therefore, another kind of process with contact terms should also be involved in our calculations \cite{Chen:2013cpa}. Their Feynman diagrams are shown in Fig.~\ref{f4}.

The Lagrangian of the contact vertexes in Fig.~\ref{f4} is written as
\begin{equation}
\mathcal{L}_{YD\bar{D}^*\gamma}=-ieg_{YD\bar{D}^*\gamma}\varepsilon^{\mu\nu\alpha\beta}A_\nu\left[D^*_{\mu}\bar{D}
-D\bar{D}^*_{\mu}\right]\partial_\beta Y_\alpha,\\
\end{equation}

\begin{equation}
\mathcal{L}_{XD\bar{D}^*\gamma}=-\frac{i}{\sqrt{2}}ef_{XD\bar{D}^*\gamma}A^\mu B^\alpha\left(D_\mu^*\overleftrightarrow{\partial}_\alpha\bar{D}-D\overleftrightarrow{\partial}_\alpha\bar{D}_\mu^*\right).
\end{equation}

With the above contact Lagrangians, the amplitudes of Feynman diagrams are

\begin{equation}
\begin{split}
\mathcal{M}^e=&\int \frac{d^4p_1}{(2\pi)^4}(-i)eg_{YD\bar{D}^*\gamma}\varepsilon^{\mu\nu\alpha\beta}\varepsilon_{3\nu}^*p_{A\beta}\varepsilon_{A\alpha}\\
&\times\left(-g_{\mu\rho}+\frac{p_{2\mu}p_{2\rho}}{m_2^2}\right)\left(\frac{g_{YD\bar{D}^*}}{\sqrt{2}}\varepsilon_4^{*\rho}\right)\\
&\times\frac{\mathcal{F}_{cont}^2(p_1^2)}{(p_1^2-m_1^2)(p_2^2-m_2^2)},\\
\end{split}
\end{equation}

\begin{equation}
\begin{split}
\mathcal{M}^f=&\int \frac{d^4p_1}{(2\pi)^4}(-i)eg_{YD\bar{D}^*\gamma}\varepsilon^{\mu\nu\alpha\beta}\varepsilon_{3\nu}^*p_{Y\beta}\varepsilon_{Y\alpha}\\
&\times\left(-g_{\mu\rho}+\frac{p_{1\mu}p_{1\rho}}{m_1^2}\right)\left(\frac{g_{YD\bar{D}^*}}{\sqrt{2}}\varepsilon_4^{*\rho}\right)\\
&\times\frac{\mathcal{F}_{cont}^2(p_1^2)}{(p_1^2-m_1^2)(p_2^2-m_2^2)},\\
\end{split}
\end{equation}

\begin{equation}
\begin{split}
\mathcal{M}^g=&\int \frac{d^4p_1}{(2\pi)^4}(-i)g_{YD\bar{D}^*}\varepsilon^{\mu\nu\alpha\beta}p_{A\beta}\varepsilon_{A\alpha}\\
&\times(p_1-p_2)_\nu\left(-g_{\mu\delta}+\frac{p_{2\mu}p_{2\delta}}{m_2^2}\right)\frac{ef_{XD\bar{D}^*\gamma}}{\sqrt{2}}\varepsilon_3^{*\delta}\varepsilon_4^{*\sigma}\\
&\times(p_2-p_1)_\sigma\frac{\mathcal{F}_{cont}^2(p_1^2)}{(p_1^2-m_1^2)(p_2^2-m_2^2)},\\
\end{split}
\end{equation}

\begin{equation}
\begin{split}
\mathcal{M}^h=&\int \frac{d^4p_1}{(2\pi)^4}(-i)g_{YD\bar{D}^*}\varepsilon^{\mu\nu\alpha\beta}p_{A\beta}\varepsilon_{A\alpha}(p_1-p_2)_\nu\\
&\times\left(-g_{\mu\rho}+\frac{p_{1\mu}p_{1\rho}}{m_1^2}\right)\frac{ef_{XD\bar{D}^*\gamma}}{\sqrt{2}}\varepsilon_3^{*\rho}\varepsilon_4^{*\sigma}\\
&\times(p_2-p_1)_\sigma\frac{\mathcal{F}_{cont}^2(p_1^2)}{(p_1^2-m_1^2)(p_2^2-m_2^2)}.\\
\end{split}
\end{equation}

In the amplitudes $\mathcal{M}^{(e\sim h)}$, an additional form factor $\mathcal{F}_{\rm cont}(p_1^2)$ is introduced to guarantee the gauge invariance of the photon field for the long-distance interaction shown in Fig.~\ref{f2} and Fig.~\ref{f4}. As a result, $\mathcal{F}_{\rm cont}(p_1^2)$ is not an independent entity; it is determined by the gauge invariance condition.

The gauge invariance of the photon field is expressed through the Ward-Takahashi Identity. We have noticed that the amplitudes $\mathcal{M}^c$ and $\mathcal{M}^d$ satisfy the Ward-Takahashi Identity directly, while the Ward-Takahashi Identity is broken in the amplitudes $\mathcal{M}^a$, $\mathcal{M}^b$, $\mathcal{M}^e$, $\mathcal{M}^f$, $\mathcal{M}^g$, and $\mathcal{M}^h$. By summing over all the amplitudes $\mathcal{M}^a\sim \mathcal{M}^h$, the total amplitude will satisfy the relation $p_3^\mu\mathcal{M}^{\rm tot}_\mu=0$, where $\mathcal{M}^{\rm tot}_\mu$ is the total amplitudes without the polarized vector of the photon $\varepsilon_3^\mu$. After dealing with loop integrals, the Lorentz structure of the amplitudes can be expanded as
\begin{equation}
\begin{split}
\mathcal{M}^a=&A_1\varepsilon_{\mu\nu\alpha\beta}p_3^\mu\varepsilon_3^\nu\varepsilon_4^\alpha\varepsilon_A^\beta+A_2\varepsilon_{\mu\nu\alpha\beta}p_4^\mu\varepsilon_3^\nu\varepsilon_4^\alpha\varepsilon_A^\beta\\
&+A_3\varepsilon_{\mu\nu\alpha\beta}p_3^\mu p_4^\nu\varepsilon_4^\alpha\varepsilon_A^\beta p_4\cdot\varepsilon_3,\\
\mathcal{M}^b=&B_1\varepsilon_{\mu\nu\alpha\beta}p_3^\mu\varepsilon_3^\nu\varepsilon_4^\alpha\varepsilon_A^\beta+B_2\varepsilon_{\mu\nu\alpha\beta}p_4^\mu\varepsilon_3^\nu\varepsilon_4^\alpha\varepsilon_A^\beta\\
&+B_3\varepsilon_{\mu\nu\alpha\beta}p_3^\mu p_4^\nu\varepsilon_3^\alpha\varepsilon_A^\beta p_3\cdot\varepsilon_4\\
&+B_4\varepsilon_{\mu\nu\alpha\beta}p_3^\mu p_4^\nu\varepsilon_4^\alpha\varepsilon_A^\beta p_4\cdot\varepsilon_3,\\
\mathcal{M}^e=&\mathcal{M}^f=E_1\varepsilon_{\mu\nu\alpha\beta}p_3^\mu\varepsilon_3^\nu\varepsilon_4^\alpha\varepsilon_A^\beta
+E_2\varepsilon_{\mu\nu\alpha\beta}p_4^\mu\varepsilon_3^\nu\varepsilon_4^\alpha\varepsilon_A^\beta,\\
\mathcal{M}^g=&\mathcal{M}^h=G_1\varepsilon_{\mu\nu\alpha\beta}p_3^\mu\varepsilon_3^\nu\varepsilon_4^\alpha\varepsilon_A^\beta
+G_2\varepsilon_{\mu\nu\alpha\beta}p_4^\mu\varepsilon_3^\nu\varepsilon_4^\alpha\varepsilon_A^\beta,\\
\end{split}
\end{equation}
where the parameters have the relations, i.e., $A_1=A_2$, $B_1=B_2$, $E_1=E_2$, and $G_1=G_2$.

In the contact amplitudes ($\mathcal{M}^e$$\sim$$\mathcal{M}^h$), only $E_1\varepsilon_{\mu\nu\alpha\beta}p_3^\mu\varepsilon_3^\nu\varepsilon_4^\alpha\varepsilon_A^\beta$ and  $G_1\varepsilon_{\mu\nu\alpha\beta}p_3^\mu\varepsilon_3^\nu\varepsilon_4^\alpha\varepsilon_A^\beta$ guarantee the gauge invariance of the photon field. Thus, the total contribution from the contact amplitudes in Fig.~\ref{f4} can be written as
\begin{equation}
\begin{split}
\mathcal{M}^{\rm cont}=(2E_1+2G_1)\varepsilon_{\mu\nu\alpha\beta}p_3^\mu\varepsilon_3^\nu\varepsilon_4^\alpha\varepsilon_A^\beta,\\
\end{split}
\end{equation}
where the parameter $2E_1+2G_1$ can be expressed by $2E_1+2G_1=-A_2-B_2+A_3(p_4\cdot p_3)+B_4(p_4\cdot p_3)$, based on the relation $p_3^\mu\mathcal{M}^{\rm tot}_\mu=0$. And only the $A_1\varepsilon_{\mu\nu\alpha\beta}p_3^\mu\varepsilon_3^\nu\varepsilon_4^\alpha\varepsilon_A^\beta$, $B_1\varepsilon_{\mu\nu\alpha\beta}p_3^\mu\varepsilon_3^\nu\varepsilon_4^\alpha\varepsilon_A^\beta$, and $B_3\varepsilon_{\mu\nu\alpha\beta}p_3^\mu p_4^\nu\varepsilon_3^\alpha\varepsilon_A^\beta p_3\cdot\varepsilon_4$ terms can guarantee the gauge invariance of photon field in $\mathcal{M}^a$ and $\mathcal{M}^b$. Therefore, combined with $\mathcal{M}^c$ and $\mathcal{M}^d$, these Ward-Takahashi Identity allowed terms in triangle diagrams of Fig.~\ref{f2} are collected as
\begin{equation}
\begin{split}
\mathcal{M}^{\rm tri}=&\mathcal{M}^c+\mathcal{M}^d+(A_1+B_1+B_3(p_3\cdot\varepsilon_4))\varepsilon_{\mu\nu\alpha\beta}p_3^\mu\varepsilon_3^\nu\varepsilon_4^\alpha\varepsilon_A^\beta.\\
\end{split}
\end{equation}

With the above treatment, the gauge invariance of the photon field has been guaranteed. The loop amplitude of $Y\to \gamma X(3872)$ process shown in the Fig.~\ref{f2} and Fig.~\ref{f4} can be written as
\begin{equation}
\begin{split}
\mathcal{M}^{\rm loop}=\mathcal{M}^{\rm cont}+\mathcal{M}^{\rm tri}.\\
\end{split}
\end{equation}

Finally, with the amplitude $\mathcal{M}^{\rm loop}$, the corresponding decay width can be expressed as
\begin{equation}
\begin{split}
\Gamma^{\rm loop}[Y(4200)\to\gamma X(3872)]=\frac{1}{3}\frac{1}{8\pi}\frac{|\textbf{p}_3|}{m_A^2}|\mathcal{M}^{\rm loop}|^2.\\
\end{split}
\end{equation}

The dependance of $\Gamma^{\rm loop}[Y(4200)\to\gamma X(3872)]$ on $\alpha$ is shown in the Table.~\ref{Tab2}, where the parameter $Z_X$ is not determined.
\begin{table}[htbp]
\caption{The long-distance contribution of the radiative decay from $\mathcal{M}^a$ to $\mathcal{M}^h$. Here, the $Y_1$ and $Y_2$ denote the $\psi(4160)$ and $\psi(4230)$, respectively.}
\label{Tab2}
\renewcommand\arraystretch{1.50}
\begin{tabular*}{86mm}{@{\extracolsep{\fill}}cccccccc}
\toprule[1.5pt]
\toprule[1pt]
$\alpha$ &1.0&1.5&2.0&2.5&3.0&3.5&4.0\\
\midrule[1pt]
$\frac{\Gamma^{\rm loop}[Y_1\to\gamma X(3872)]}{1-Z_X}$ (eV)&0.7&1.1&1.6&2.3&3.2&4.5&5.7\\
$\frac{\Gamma^{\rm loop}[Y_2\to\gamma X(3872)]}{1-Z_X}$ (eV)&0.4&0.7&1.0&1.5&1.9&2.9&3.8\\
\bottomrule[1pt]
\bottomrule[1.5pt]
\end{tabular*}
\end{table}

\begin{center}
\begin{figure}[htbp]
\includegraphics[width=5.0cm,keepaspectratio]{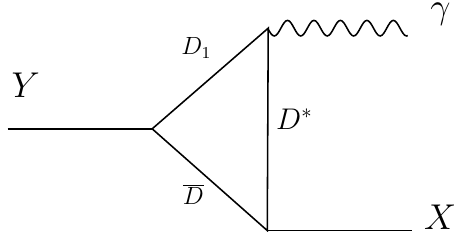}
\caption{The long-distance radiative decay through $D_1\bar{D}$ channel.}
\label{fDD1}
\end{figure}
\end{center}

The threshold of $D_1\bar{D}$ channel is around 4.289 GeV ($m_D=$1.867 GeV and $m_{D_1}=$2.422 GeV), which is close to 4.2 GeV. Meanwhile, the $D_1\bar{D}$ pair could form a molecular state with $J^{P}=1^{-}$ \cite{Ding:2008gr, Qin:2016spb}, which should be included in the $Y(4200)$ structure. The $D_1\bar{D}$ can naturally couple to $J^{PC}=1^{--}$ through an $S$-wave interaction, and the $D_1D^*\gamma$ vertex is an $S$-wave interaction as well. Therefore the radiative decay process in Fig.~\ref{fDD1} through $D_1\bar{D}$ channel ought to be significant, which should be considered in our calculations.

The vertexes of $YD_1\bar{D}$ coupling and $D_1D^*\gamma$ coupling in Fig.~\ref{fDD1} are depicted by the Lagrangians, i.e.,
\begin{equation}\label{eqDD1}
\begin{split}
\mathcal{L}_{YDD_1}&=\frac{g_Y}{\sqrt{2}}Y_\mu(D_1^\mu\bar{D}-D\bar{D}_1^\mu),\\
\mathcal{L}_{D_1D^*\gamma}&=ieg_{D_1D^*\gamma}\varepsilon_{\mu\nu\alpha\beta}F^{\mu\nu}D^{*\alpha\dag} D_1^\beta.\\
\end{split}
\end{equation}
With the Lagrangian in Eq.~(\ref{eqDD1}), the amplitude of the process in Fig.~\ref{fDD1} can be expressed as
\begin{equation}
\begin{split}
\mathcal{M}^{D\bar{D}_1}=&\int \frac{d^4q}{(2\pi)^4}(-i)\frac{g_Y}{\sqrt{2}}\varepsilon_{A\rho}(-g^{\rho\beta}+\frac{p_1^\rho p_1^\beta}{m_1^2})\\
&\times eg_{D_1D^*\gamma}\varepsilon_{\mu\nu\alpha\beta}(p_3^\mu\varepsilon_3^{*\nu}-p_3^\nu\varepsilon_3^{*\mu})
\frac{g_X}{\sqrt{2}}\varepsilon_{4\tau}^*\\
&\times (-g^{\alpha\tau}+\frac{q^\alpha q^\tau}{m_q^2})\frac{\mathcal{F}^2(q^2)}{(p_1^2-m_1^2)(q^2-m_q^2)(p_2^2-m_2^2)},\\
\end{split}
\end{equation}
where the coupling constant of $D_1D^*\gamma$ vertex is determined as $g_{D_1D^*\gamma}=0.875$ by fitting the calculated width
$\Gamma_{D_{1} \to D^{*} \gamma} = 166~{\rm keV}$ from Ref.~\cite{Korner:1992pz}. $g_{Y}$ is the physical coupling constant for the molecular state $Y$ with its component $D_1\bar{D}$, which was fixed through the study on $e^+e^-\to Y(4260)\to D\bar{D}^*\pi$ process in Ref.~\cite{Qin:2016spb}, therefore we employed the coupling constant $g_{Y}=3.94~{\rm GeV}^{-1/2}$.

Finally, with the above amplitude the decay width through $D_1\bar{D}$ channel is calculated. Combining the $\mathcal{M}^{\rm loop}$ and $\mathcal{M}^{D_1\bar{D}}$, we can get the total amplitude for the long-distance part of $Y(4200)\to \gamma X(3872)$ process. We calculate the decay widths from the total amplitude of long-distance interactions and present them in Table~\ref{Tab3}.

\begin{table}[htbp]
\caption{The radiative decay widths from a molecular $D_1\bar{D}$ initial state and the total radiative decay widths $\Gamma^L$ in the long-distance interaction. Here, the $Y_1$ and $Y_2$ represent the $\psi(4160)$ and $\psi(4230)$, respectively.}
\label{Tab3}
\renewcommand\arraystretch{1.50}
\begin{tabular*}{86mm}{@{\extracolsep{\fill}}cccccccc}
\toprule[1.5pt]
\toprule[1pt]
$\alpha$ &1.0&1.5&2.0&2.5&3.0&3.5&4.0\\
\midrule[1pt]
$\frac{\Gamma^{D_1\bar{D}}[Y_1\to\gamma X(3872)]}{1-Z_X}$ (eV)&4.7&6.6&8.1&9.4&10.6&11.6&12.5\\
$\frac{\Gamma^{L}[Y_1\to\gamma X(3872)]}{1-Z_X}$ (eV)&5.0&7.5&10.1&13.4&16.7&20.7&24.1\\
\midrule[1pt]
$\frac{\Gamma^{D_1\bar{D}}[Y_2\to\gamma X(3872)]}{1-Z_X}$ (eV)&4.9&6.8&8.4&9.7&10.8&11.9&12.8\\
$\frac{\Gamma^{L}[Y_2\to\gamma X(3872)]}{1-Z_X}$ (eV)&7.5&10.8&14.1&17.8&21.1&25.8&29.8\\
\bottomrule[1pt]
\bottomrule[1.5pt]
\end{tabular*}
\end{table}

Since only the long-distance interaction is included in the $\Gamma^{L}[Y\to\gamma X(3872)]$, the coupling constant $g_{YX\gamma}^L$ can be estimated by $\Gamma^{L}[Y_{1(2)}\to\gamma X(3872)]/(1-Z_X)=24.1~(29.8)~{\rm eV}$. Here, by using the larger results obtained with $\alpha=4$, we artificially overestimate the decay width in order to outstand the contribution of the long-distance interaction sufficiently.
As a result, the coupling constant is determined as $g_{YX\gamma}^L=0.026\sqrt{1-Z_X}$.\footnote{The two decay widths $\Gamma^{L}[Y_{1(2)}\to\gamma X(3872)]/(1-Z_X)=24.1~(29.8)~{\rm eV}$ will give two similar $g_{YX\gamma}^L$s around $0.026\sqrt{1-Z_X}$. Their tiny differences can be ignored in our study.}

Except for the $Z_X$, the coupling constants of the total amplitude in Eq.~(\ref{eqamptot}) are all fixed through the above calculations. Then, the total cross section can be calculated with the equation
\begin{equation}
d\sigma=\frac{1}{8(2\pi)^4}\frac{1}{4|\mathbf{k}_e|\sqrt{s}}|\mathcal{M}^{tot}|^2dk_5^0dk_3^0d\cos\theta d\eta,
\end{equation}
where $k_3^0$ and $k_5^0$ are the energy of the $J/\psi$ and $\gamma$ in the final states. In the next section, we will show the numerical results of the cross section.

\section{results and discussions}\label{sec3}

\begin{center}
\begin{figure}[htbp]
\includegraphics[width=8.6cm,keepaspectratio]{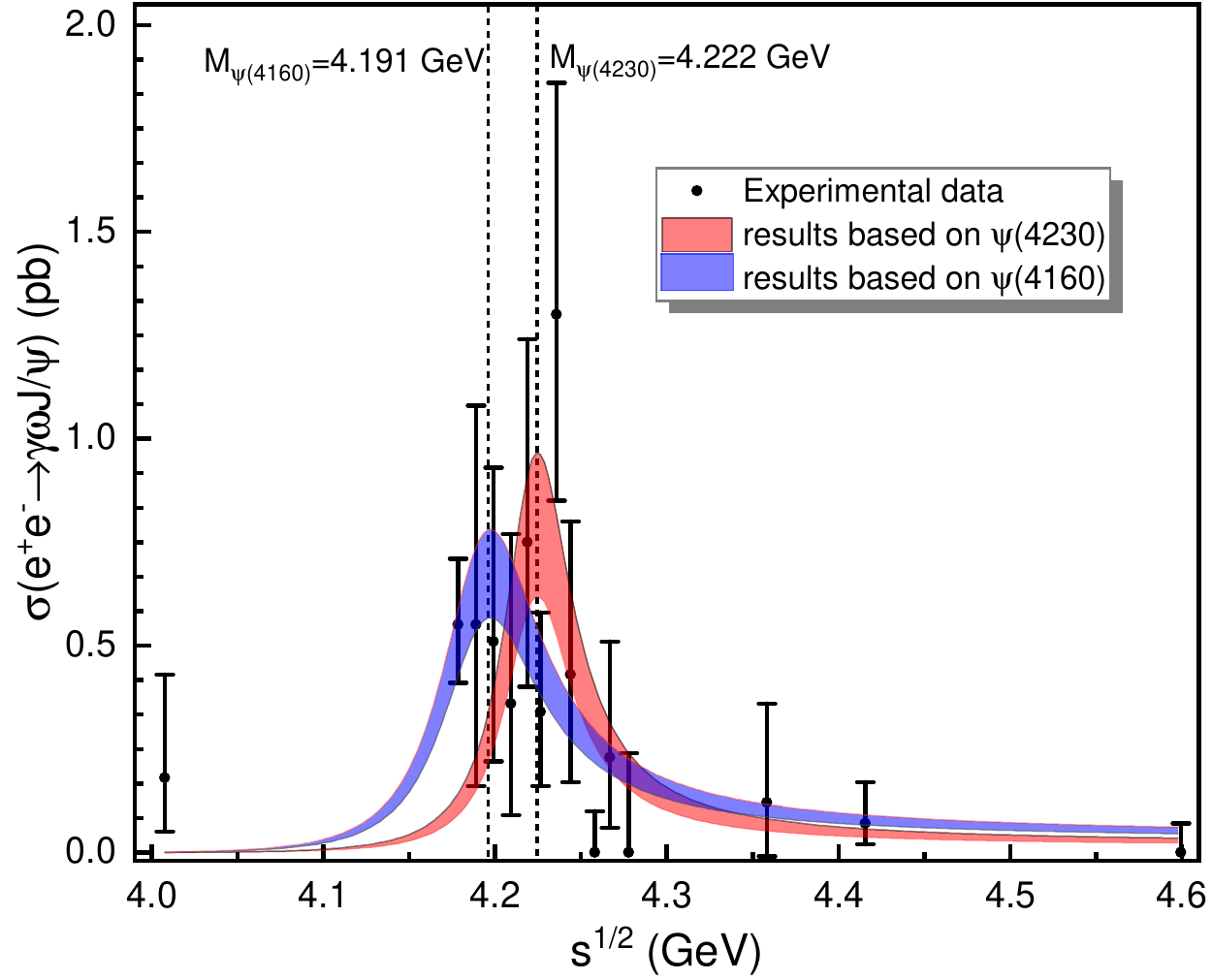}
\caption{The cross section of $e^+e^-\to\gamma J/\psi\omega$. The black dots with error bars are the experimental data copied from Ref.~\cite{BESIII:2019qvy}. The two peaks with colored bands are the numerical results when $\psi(4160)$ and $\psi(4230)$ are assumed to be the original source of $Y(4200)$.}
\label{fres}
\end{figure}
\end{center}

At this stage, the remaining undetermined variable is $Z_X$, defined in Eq.~(\ref{eq3872}). Theoretically, the value of $Z_X$ is restricted in the range [0, 1] to denote the fraction of the $c\bar{c}$ component in the $X(3872)$. For numerical analysis, the value of $Z_X$ will be deduced by comparing our calculated results for $\sigma[e^+e^-\to \gamma X(3872)\to \gamma \omega J/\psi]$ with the experimental data in Ref.~\cite{BESIII:2019qvy}. We illustrate both the theoretical and experimental results in Fig.~\ref{fres}.

Fig.~\ref{fres} presents data points with error bars, which are collected from experimental data in the supplemental material of Ref.~\cite{BESIII:2019qvy}. In the experiment, the $X(3872)$ particle is identified via the $J/\psi \omega$ decay channel, which enable the measurement of the cross section for the process $e^+e^-\to \gamma X(3872)\to \gamma\omega J/\psi$. To determine the most accurate value of $Z_X$ based on the data depicted in Fig.~\ref{fres}, we employ a minimization approach, defining $\chi^2=\sum_i(\sigma_i^{\rm exp}-\sigma_i^{\rm cal})^2$ as the error function to quantify the discrepancies between experimental and theoretical cross sections. With this method, we can compare the numerical predictions for various $Z_X$ values against experimental data directly.

If assigning the $\psi(4160)$ resonance as the source of $Y(4200)$ structure, we yield a $Z_X$ value in the range of 0.25 to 0.28, with the minimized $\chi^2$ value at $1.2^{+1.2}_{-0.9}$. This scenario is illustrated by the left peak highlighted with a blue band in Fig.~\ref{fres}. Conversely, when the $\psi(4230)$ resonance is used to describe the $Y(4200)$ structure, $Z_X$ is determined between 0.12 and 0.14, with the minimized $\chi^2$ at $1.3^{+1.0}_{-0.6}$. This finding is illustrated by the right peak, marked with a red band in Fig.~\ref{fres}.

Since the E1 decay width $\Gamma^{E1}[\psi(2D)\to\gamma \chi_{c1}(2P)]$ is significantly larger than $\Gamma^{L}[Y(4200)\to\gamma X(3872)]$, the short-distance coupling strength is far stronger than the strength of the long-distance coupling. As a result, if only the long-distance interaction is involved in the calculation, we are difficult to explain the experimental cross section data in Fig.~\ref{fres}. By adjusting the value of $Z_X$, the short-distance contribution of $Y(4200)\to \gamma X(3872)$ process is involved. We find the magnitudes and lineshapes of our numerical results can explain the experimental data points of $\sigma[e^+e^-\to \gamma \omega J/\psi]$ successfully, after the short-distance contribution is considered. In detail from the study, we arrive at the following four conclusions:
\begin{itemize}

\item The short-distance and long-distance interactions exist in the $e^+e^-\to \gamma X(3872)$ process simultaneously. Comparing to the long-distance interaction, the strength of the short-distance interaction induced by the E1 transition is much larger.

\item  Although the strength of the short-distance interaction is much larger in the radiative production of $X(3872)$, the $Z_X$ value derived from the experimental data suggests that there is only a 12\% to 28\% likelihood of identifying an $X(3872)$ as a short-distance $\chi_{c1}(2P)$ state. The results imply that the $X(3872)$ is still mainly composed of $D\bar{D}^*$ molecular component.

\item Comparing the $\Gamma^{D_1\bar{D}}[Y\to\gamma X(3872)]$ and $\Gamma^{\rm loop}[Y\to\gamma X(3872)]$, we find the $S-$wave interaction of the molecular $D_1\bar{D}$ pair will significantly enhance the contribution to the long-distance radiative decay process. However, the contribution remains much smaller than that of the leading-order E1 transition.

\item Based on the numerical results, the present experimental data are insufficient to clarify which state the $Y(4200)$ structure originates from $\psi(4160)$ or $\psi(4230)$. Consequently, more detailed measurement of the process is necessary.

\end{itemize}

Besides the above numerical calculations, we also carry out some simple estimation with the remaining experimental information. In Ref.~\cite{BESIII:2019qvy}, the relative branch ratio $\mathcal{R}$ and $\Gamma^{e^+e^-\to \gamma J/\psi\pi\pi}_{Y(4200)}$ are measured directly, where $\mathcal{R}$ is defined as $\mathcal{R}=\mathcal{B}[X(3872)\to\omega J/\psi]/\mathcal{B}[X(3872)\to\pi^+\pi^-J/\psi]$, and $\Gamma^{e^+e^-\to \gamma J/\psi\pi\pi}_{Y(4200)}=\Gamma^{e^+e^-}\times\mathcal{B}[Y(4200)\to \gamma X(3872)]\times\mathcal{B}[X(3872)\to \pi^+\pi^- J/\psi]$. With above definitions, the decay width of $Y(4200)\to \gamma X(3872)$ can be estimated by
\begin{equation}
\begin{split}
\Gamma^{\rm est}[Y(4200)\to\gamma X(3872)]=\frac{\mathcal{R}\cdot\Gamma^{e^+e^-\to \gamma J/\psi\pi^+\pi^-}_{Y(4200)}\cdot\Gamma_{Y(4200)}^{\rm tot}}{\Gamma^{e^+e^-}_{Y(4200)}\cdot\mathcal{B}[X(3872)\to\omega J/\psi]},
\end{split}
\end{equation}
where $\mathcal{R}=1.6^{+0.4}_{-0.3}$, $\Gamma^{e^+e^-\to \gamma J/\psi\pi^+\pi^-}_{Y(4200)}=4.5^{+1.1}_{-0.8}\times10^{-2}~{\rm eV}$ and $\mathcal{B}[X(3872)\to\omega J/\psi]=4.3\pm2.1 \%$ are determined by experiments directly \cite{ParticleDataGroup:2022pth}. Here, $\Gamma^{e^+e^-}_{\psi(4160)}=0.48\pm0.22~{\rm keV}$, $\Gamma^{\rm tot}_{\psi(4160)}=70\pm10~{\rm MeV}$ and $\Gamma^{\rm tot}_{\psi(4230)}=48\pm8~{\rm MeV}$, $\Gamma^{e^+e^-}_{\psi(4230)}\simeq\Gamma^{\mu^+\mu^-}_{\psi(4230)}=1.53\pm1.26\pm0.54~{\rm keV}$ are employed separately to roughly evaluate $\Gamma^{\rm est}[Y_{1(2)}\to\gamma X(3872)]$. The results can be obtained, i.e.,
\begin{equation}
\begin{split}
\Gamma^{\rm est}[Y_1\to\gamma X(3872)]&=244^{+188}_{-179}~{\rm keV},\\
\Gamma^{\rm est}[Y_2\to\gamma X(3872)]&=53^{+57}_{-56}~{\rm keV}.\\
\end{split}
\end{equation}
Through the above estimation, the present experimental data from either $\psi(4160)$ or $\psi(4230)$ all suggest that $\Gamma^{\rm est}[Y(4200)\to\gamma X(3872)]$ falls within the order about 10$\sim$100~keV, which is in accordance with the E1 transition result in the Table~\ref{Tab1} and far different from the decay widths induced by the long-distance interaction shown in Tables~\ref{Tab2} and \ref{Tab3}. The above estimations also suggest that the cross-section data of $e^+e^-\to \gamma X(3872)$ process are dominated by the short-distance interaction between $c\bar{c}$ components.

Besides the $\psi(2D)\to \chi_{c1}(2P)\gamma$ process listed in the Table~\ref{Tab1}, the $\psi(3S)\to \chi_{c1}(2P)\gamma$ transition also exhibits a notable decay width in the E1 radiative transition. Therefore, the resonance signals belong to $\psi(4040)$ should exist around 4.040 GeV in the cross-section data. However, as depicted in Fig.~\ref{fres}, there is only a single experimental data point below the 4.15 GeV, which is insufficient for a comprehensive analysis around 4.0 GeV region. Specifically, the solitary data point at 4.0076 GeV cannot match with the current cross-section analysis of the $Y(4200)$. But if the $\psi(4040)$ is involved in the $\sigma[e^+e^-\to \gamma X(3872)]$, it could be understood directly. Nevertheless, without additional data collected between 4.0076 GeV and 4.1783 GeV, we are unable to make a further study on whether the $\psi(4040)$ peak exists in the $\sigma[e^+e^-\to \gamma X(3872)]$.
\section{summary}\label{sec4}

After the $X(3872)$ is found by the Belle Collaboration, theorists pay much efforts to make clear the nature of the $X(3872)$. Along with the progress of the different experiments, experimentalists accumulated a plenty of data of the $X(3872)$ across different processes, which supports the deep theoretical research to its structure. Focusing on the cross-section data of the $Y(4200)\to \gamma X(3872)$, we study the long-distance and short-distance interaction in this process through calculating the E1 radiative transition and hadronic loop diagrams.

Comparing with the experimental results, the E1 transition makes $\sigma[e^+e^-\to\gamma X(3872)\to\gamma\omega J/\psi]$ possible to reach the measured cross-section data, which clearly reveals the short-distance interaction is crucial in the process. And the compact $c\bar{c}$ components are necessary to understand the structure of $X(3872)$ and $Y(4200)$.

Through this study, we notice that the present mass and width for $Y(4200)$ still have large uncertainties in the experiment, which should be promoted in the future. Besides, the missing cross section between 4.0076 GeV and 4.1783 GeV in the experiment may be noteworthy, since the potential presence of the $\psi(4040)$ peak within this interval will help us understand the $Y(4200)$ more clearly. We hope that future experiments could obtain more comprehensive and precise data, which will enhance our understanding of this process and the relative topics.
\section*{Acknowledgements}
The author is grateful to acknowledge Prof. Dian-Yong Chen and Prof. Qiang Zhao for helpful suggestions and discussions. This work is supported by the National Natural Science foundation of China (Grant No.12347135) and China Postdoctoral Science Foundation No. 2023M733502

\end{document}